\documentclass[a4paper,11pt]{article}
\usepackage{jcappub} 
\usepackage{lineno}
\newcommand{\smpc}{\,h^{-1}\mathrm{Mpc}}
\newcommand{\kmpc}{\,\mathrm{Mpc}^{-1}h}

\title{\boldmath The Poisson noise in modeling the redshift-space distortion at large scales}

\author[a,b]{HongXiang Chen}
\affiliation[a]{National Astronomical Observatories, Chinese Academy of Sciences\\
Beĳing, 100012, China}
\affiliation[b]{School of Astronomy and Space Science, University of Chinese Academy of Sciences\\
Beĳing 100049, China}

\author[a,b]{Jie Wang}

\author[c]{Baojiu Li}
\affiliation[c]{Institute for Computational Cosmology, Department of Physics, Durham University\\
Durham DH1 3LE, UK}
\emailAdd{chxiang@nao.cas.cn, jie.wang@nao.cas.cn, baojiu.li@durham.ac.uk}

\abstract{
We investigate the errors in modeling the redshift-space distortion (RSD) effect at large linear scales, using data from the Millennium simulation. While standard theoretical templates, such as the Kaiser formula and the TNS method, could precisely model RSD for individual large-scale modes, we find that for tracers with number densities lower than $\sim10^{-3}({\rm Mpc}/h)^{-3}$, there is a few-percent level bias in the predicted power spectrum. This error arises due to the amplification of intrinsic Poisson noise during RSD modeling from real-space power spectrum. This amplified noise can be analytically expressed as $1 + \epsilon/[{\bar{n}P}({1+\epsilon})]$, with $\epsilon=2\beta/3+\beta^2/5$, where $P$ denotes the real-space tracer power spectrum and $\beta \equiv f/b$. Specifically, for halos with a number density of around $5\times10^{-4}({\rm Mpc}/h)^{-3}$, this phenomenon results in an additional systematic error of 2.5\%. Our result suggests that caution is necessary when directly modeling redshift-space distortions (RSD) using real-space power spectra of tracers obtained from simulations or actual surveys. This caution is particularly pertinent in scenarios where emulators trained on simulation data forecast the real-space tracer power spectrum, as well as in baryon acoustic oscillation (BAO) reconstruction using galaxy samples, for which we estimate that shot noise could introduce random errors of about one-third in the displacement field, potentially diminishing the effectiveness of the BAO peak sharpening.}


\begin{document}
\maketitle
\flushbottom

\section{Introduction} \label{sec:intro}

In recent years, large galaxy redshift surveys such as BOSS \citep{Alam2017}, eBOSS \citep{Alam2021}, DESI \citep{DESICollaboration2016} and Euclid \citep{Laureijs2011} have been contributing to constraints on the parameters of the standard $\Lambda$ cold dark matter ($\Lambda$CDM) model at the percent level. Achieving this level of precision in parameter constraints requires a reevaluation of the underlying systematic errors, including on larger scales where theoretical predictions are considered to be accurate. 

A key cosmological probe relevant to these galaxy surveys is galaxy clustering. In the study of galaxy clustering, as the positions of galaxies along the line of sight are influenced by the redshift-space distortion (RSD) effect \citep{Peebles1980}, it is crucial to accurately account for this effect in our theoretical predictions in order to reliably use observational data to test the cosmological model. 

The Kaiser formula, introduced by \citep{Kaiser1987}, offers the first robust quantitative description of the large-scale RSD effect, drawing from linear perturbation theory for the cosmic density and velocity fields. Following this early work, advancements in RSD modeling have extended from linear to quasi-linear and nonlinear domains. These progresses have been achieved through the application of higher-order perturbation theory and the halo model \citep{Croft1997,Nusser1991,Branchini1999,Nusser1994,Fisher1995,kudlicki2000, Branchini2002, Landy2002,Scoccimarro2004, Erdogdu2006, Taruya2010, Kitaura2012,Zhang2013,Tanimura2022,GaneshaiahVeena2022}, iterative methods \citep{Yahil1991, Wang2009,Kitaura2012, Wang2012, Shi2016,  Wang2020}, and artificial neural networks \citep{Chen2023, Wu2023, Qin2023}.

Despite significant efforts directed toward refining the RSD model for smaller scales, the accuracy in modeling RSD at large scales is often ignored, because linear theory works so well on those scales. However, depending on how the ingredients of the RSD model are obtained, errors at a few percent level could still arise on large linear scales, as will be exemplified below. In the context of precision cosmology, where percent-level accuracy in model predictions is needed, such errors can be significant, thus necessitating further investigation. The evaluation of such errors and exploration of their origin are the primary focus of this paper. 

We stress that the large-scale error to be discussed in this work only arise in certain methods of RSD modeling. Usually, the ingredients of RSD models, such as the real-space power spectrum of a tracer, are obtained from perturbation theory. However, in recent years, growing effort has been shifted to using the power spectra from simulations, either direct measurements or predictions by trained emulators. Emulator can directly interpolate statistics measured from simulations in a given parameter domain, thereby avoiding assumption about a specific functional form for parameter dependence. Some emulators employ Gaussian processes or neural networks to learn the real-space power spectrum of tracers from simulations \citep{Heitmann2009, Heitmann2014, Lawrence2017, Euclid2019,Donald2022,Chen2025}. Others directly measure the redshift-space power spectrum from simulations and use these measurements as training data to construct emulators, usually through Gaussian processes or neural networks \citep{Kobayashi2020,Wang2023}.

In this work, we focus on RSD predictions in $k$ space, where the real-space power spectra are measured from mock tracer (galaxies or dark matter halos) catalogs. To quantify the large-scale error in RSD modeling, we compare the redshift-space power spectra measured from simulation data with the predictions of two widely-used RSD models: the traditional Kaiser formula, known for its effectiveness in modeling linear RSD effects, and the TNS model (Taruya, Nishimichi \& Saito (2010)\citep*{Taruya2010}), which has extended validity from linear to mildly nonlinear scales.

This paper is organized as follows. In Section \ref{sec:theory} we describe the RSD models and the simulations used in our study. This is followed by a detailed analysis of the model errors compared with the simulation results in Section \ref{sec:result}. Finally, discussion and conclusions are presented in Section \ref{sec:conclusion}. 

\section{Models and Data}
\label{sec:theory}

In this section, we will very briefly describe the two models for large-scaleRSD: the Kaiser model and a non-linear model, TNS. Following this, we will present the simulation data employed in this paper.

\subsection{The Kaiser model}
\label{sec:Kaiser linear model} 


In the Kaiser model, the power spectrum in redshift space, $P^{(\textrm{S})}(k)$, can be written as
\begin{equation}
P^{\rm (S)}_{\rm Kaiser}(k, \mu)=P^{\rm (R)}(k)\left(1+\beta \mu^2\right)^2,  \quad 
\beta=f/b
\label{eq:1}
\end{equation} 
where $P^{\rm (R)}(k)$ is the real space power spectrum of the biased tracers, $\mu$ is the cosine of the angle between the line of sight and the line connecting the pair of galaxies, 
$f(a)={\rm d}\ln D / {\rm d} \ln a \simeq \Omega_{\mathrm{m}}^{0.6}+\frac{1}{70} \Omega_{\Lambda}\left(1+\Omega_{\mathrm{m}} / 2\right)$ \citep[e.g.][]{lahav1991} is the linear growth rate with $a,D$ being respectively the scale factor and linear growth factor and $\Omega_{\rm m},\Omega_\Lambda$ the matter and cosmological constant density parameters, and $b$ is the bias parameter, which on large linear scales can be assumed as a constant of proportionality between biased tracer (such as galaxy) and matter density fluctuations. We note that $P^{\rm (R)}(k)$ is the power spectrum of the tracer that has bias $b$. 

In this work, 
we determine the bias parameter by taking the square root of the average of ratio between the power spectra of tracers (such as halos) and of dark matter in the 9 lowest $k$-modes (where $k<0.04\kmpc$). All
power spectra are shot-noise subtracted \citep[e.g.,][]{Tinker2010}. 

\subsection{The TNS model}
\label{sec:Kaiser formula} 

While the Kaiser formula is straightforward and user-friendly, it only works for large linear scales. To accurately model the RSD at smaller scales, we also explore the TNS model \citep*{Taruya2010} which is widely used in actual galaxy surveys \citep[e.g.,][]{Oka2014,Beutler2014,Gil2016,Grieb2017}. The TNS model was developed from earlier works such as \cite{Scoccimarro2004,Percival2009}, by incorporating the one-loop perturbation for smaller scales. In this model, the redshift-space power spectrum of a generic biased tracer is given by
\begin{equation}
\begin{gathered}
P_{\textrm{TNS}}^{(\mathrm{S})}(k, \mu)=D_{\mathrm{FoG}}\left[k \mu f \sigma_{\mathrm{v}}\right]\left\{P_{\delta \delta}(k)+2 f \mu^2 P_{\delta \theta}(k)\right. \\
\left.+f^2 \mu^4 P_{\theta \theta}(k)+A(k, \mu)+B(k, \mu)\right\} .
\end{gathered}
\label{eq:TNS}
\end{equation}
where $\delta$ is the real-space density contrast of the tracers under consideration, $\theta$ the dimensionless velocity divergence $\theta \equiv -\boldsymbol{\nabla}\cdot\boldsymbol{v} /(a H f)$, $P_{ij}$ ($i,j=\delta,\theta$) the auto/cross power spectra of fields $i$ and $j$, $D_{\mathrm{FoG}}\left[k \mu f \sigma_{\mathrm{v}}\right]$ a damping function which accounts for the finger of God (FoG) effect, and the functions $A(k,\mu), B(k,\mu)$ are given as
\begin{equation}
\begin{aligned}
A(k, \mu) &=(k \mu f) \int \frac{d^3 \boldsymbol{p}}{(2 \pi)^3} \frac{p_z}{p^2} \\
& \times\left\{B_\sigma(\boldsymbol{p}, \boldsymbol{k}-\boldsymbol{p},-\boldsymbol{k})-B_\sigma(\boldsymbol{p}, \boldsymbol{k},-\boldsymbol{k}-\boldsymbol{p})\right\},\\
B(k, \mu) &=(k \mu f)^2 \int \frac{d^{3} \boldsymbol {p}}{(2 \pi)^3} F(\boldsymbol{p}) F(\boldsymbol{k}-\boldsymbol{p}),
\end{aligned}
\end{equation}
where $B_\sigma$ is defined as
\begin{equation}
\begin{gathered}
\left\langle\theta\left(\boldsymbol{k}_1\right)\left\{\delta\left(\boldsymbol{k}_2\right)+f \frac{k_{2 z}^2}{k_2^2} \theta\left(\boldsymbol{k}_2\right)\right\}\left\{\delta\left(\boldsymbol{k}_3\right)+f \frac{k_{3 z}^2}{k_3^2} \theta\left(\boldsymbol{k}_3\right)\right\}\right\rangle \\
=(2 \pi)^3 \delta_D\left(\boldsymbol{k}_1+\boldsymbol{k}_2+\boldsymbol{k}_3\right) B_\sigma\left(\boldsymbol{k}_1, \boldsymbol{k}_2, \boldsymbol{k}_3\right),
\end{gathered}
\end{equation}
and $F$ is defined as
\begin{equation}
    F(\boldsymbol{p}) \equiv \frac{p_z}{p^2}\left\{P_{\delta \theta}(p)+f \frac{p_z^2}{p^2} P_{\theta \theta}(p)\right\}.
\end{equation}
In the above $p_z$ is the $z$-component of $\boldsymbol{p}$.

In this work, $P_{\rm TNS}^{(\mathrm{S})}$ is the power spectrum of tracer in redshift space. $\delta$ in Eq.~\eqref{eq:TNS} refers to the tracers' density field in real sapce and the $\theta$ is the divergence of velocity field, which is derived from standard perturbation theory, involving the first to third order perturbation of the linear density field \citep{Heavens1998}. It should be noted that since the tracer density field is employed, the bias parameter $b$ is inherently embedded in Eq.~\eqref{eq:TNS} and the linear growth rate $f$ retains its identity, different from $\beta$ here.

The TNS model includes the corrections from the non-linear coupling between velocity and density fields, making it capable of modeling the power spectrum in redshift space for $k <  0.3 \kmpc$ and anisotropic BAO.

\subsection{Data}

The data we used to test these models is from the Millennium Simulation \citep{Springel2005}, a classical N-body simulation that assumes a $\Lambda$CDM cosmology with parameters based on a combined analysis of 2dFGRS \citep{Colless2001} and the first year WMAP data \citep{Spergel2003}:
\begin{equation}
    \left\{\Omega_{\mathrm{m}}, \Omega_{\Lambda}, \Omega_{\mathrm{b}}, h, \sigma_{8}, n_{\mathrm{s}}\right\} = \left\{ 0.25,0.75,0.045, 0.73,0.9,1.0\right\}.
\end{equation}
This cosmology is substantially different from the consensus today, but for our purpose this difference is irrelevant.

The Millennium simulation employs $2160^3$ dark matter particles in a periodic cubic box of comoving size 500 $\smpc$. The mass resolution is $8.61 \times 10^{8} h^{-1} \mathrm{M}_{\odot}$.

In observations and simulations, we usually deal with biased tracers of the underlying matter field, such as galaxies and dark matter halos. To understand the effect of different tracers in RSD modeling, we examine RSD for dark matter particles and dark matter halos at $z=0$. 
The tracer's density field ($\delta_{\rm dm}$) is calculated by cloud-in-cell (CIC) \citep{Efstathiou1985} mass assignment, using a grid of $512^3$ cells. We then use this density field to calculate the power spectrum, velocity field and bias.
The halos and subhalos are detected using the friends-of-friends (FOF) \citep{Davis1985}) and \textsc{Subfind} \citep{Springel2001} algorithms, respectively. 

\section{Results}
\label{sec:result}

In this section, we will assess the models by contrasting their predictions with simulation measurements. We first discuss the Poisson noise in the model and derive a formula to quantify the amplification of Poisson noise in RSD modeling. We then investigate the errors of different tracers with varying number densities in simulation data, confirming that the Poisson noise is amplified in the RSD modeling. Finally, we provide specific recommendations for modeling redshift-space power spectrum from real-space tracer power spectrum.


\subsection{The large-scale error in RSD Modeling}
\label{sec:Amplification of Poisson Noise}
We quantify the discrepancy between the models and the simulation measurements by the ratio of the redshift space power spectra obtained in these two approaches, $P^{\rm (S)}_{\rm model}(k)$ and $P^{\rm (S)}_{\rm sim}(k)$:
\begin{equation}
\Delta(k)= P^{\rm (S)}_{\rm model}(k) / P^{\rm (S)}_{\rm sim}(k).
\label{eq:8}
\end{equation} 
For the Kaiser and TNS models, this ratio can be expressed respectively as
\begin{equation}
\Delta_{\rm Kaiser}(k)= P^{\rm (S)}_{\rm Kaiser}(k) / P^{\rm (S)}_{\rm sim}(k),
\label{eq:9}
\end{equation} 
\begin{equation}
\Delta_{\rm TNS}(k)= P^{\rm (S)}_{\rm TNS}(k) / P^{\rm (S)}_{\rm sim}(k).
\label{eq:10}
\end{equation} 
All power spectra above are the redshift-space monopole for the tracer type considered. Here $P^{\rm (S)}_{\rm sim}(k)$ is measured from the redshift-space simulation data. The monopole is obtained by azimuthally averaging the results across all angular directions for the same $k$-mode. The redshift space tracer catalog is constructed by adding the tracers' line-of-sight velocity contribution, $v_{\rm los}/H$, to their line-of-sight positions from the simulation. The $\delta^{(\rm S)}$ field is then calculated by using the cloud-in-cell (CIC) method \citep{Efstathiou1985}, followed by Fourier transformation to find the redshift-space power spectrum. We use the same grid size of $512^3$ to calculate $\delta^{({\rm S})}$.

To accurately estimate large-scale errors, we avoid averaging the power spectra in redshift and real space with wide $k$-mode bins. Instead, we make a direct mode-by-mode comparison for the $k$-modes at the largest scales. 

\begin{figure*}
	\includegraphics[width=\textwidth] {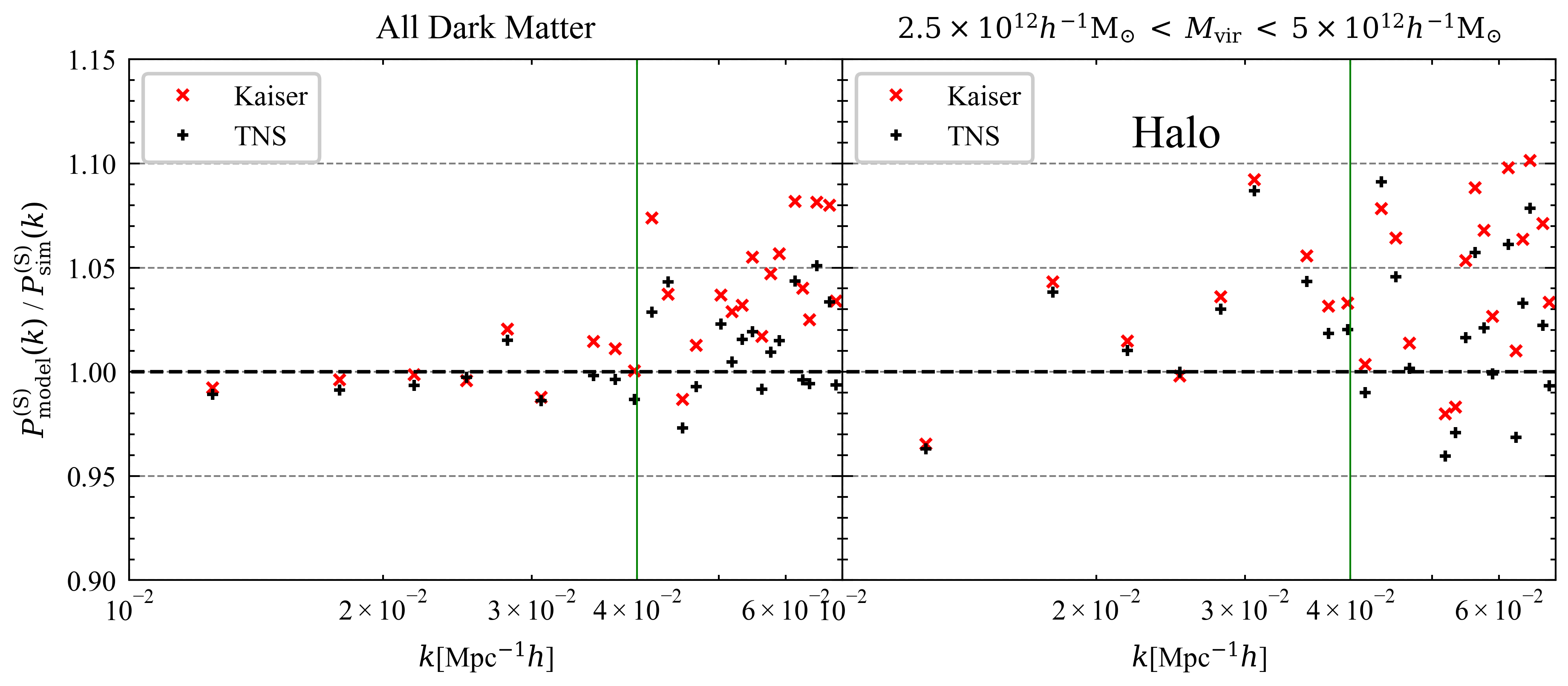}
    \caption{The largest-scale errors, defined in Eqs.~(\ref{eq:9},\ref{eq:10}), in the redshift-space power spectra of dark matter (left panel) and halos in the mass range of $M/\left(h^{-1} \mathrm{M}_{\odot}\right)\in[2.5\times10^{12},5\times10^{12}]$ (right panel). 
    The red crosses and black pluses indicate the Kaiser and TNS  models, respectively. The vertical line indicate the scales range $k < 0.04 \kmpc$ we adopt to calculate the halo bias parameter and to consider the model error.} 
    \label{fig:Kaiser1}
\end{figure*}

Figure~\ref{fig:Kaiser1} shows this result for two tracers: dark matter particles and halos within mass range $M/\left(h^{-1} \mathrm{M}_{\odot}\right)\in\left[2.5 \times 10^{12}, 5.0 \times 10^{12}\right]$, in the left and right panels respectively. The red crosses and black pluses are for the Kaiser formula, Eq.~(\ref{eq:1}), and the TNS model, Eq.~(\ref{eq:TNS}). For dark matter particles, the errors in both models are minimal, below $2\%$ for large scales ($k < 0.04 \kmpc$, which is indicated by a green vertical line), but increase sharply beyond this point, exceeding $5\%$. For dark halos, both models show larger errors, mostly surpassing $2\%$ even at $k < 0.04\kmpc$. 

In particular, for the scales which we focus on, $k < 0.04 \kmpc$, the disparity between the two models is negligible, generally much less than $1\%$. For smaller-scale modes, their disparity increases significantly, which is anticipated because the TNS model incorporates nonlinear perturbations.

Next we show that the difference in the model performance when applied to dark matter and halos has arisen from the much lower number density of dark matter halos, $\bar{n}_{\rm h}$, which introduces signifiant Poisson shot noise in the real-space power spectrum that is further amplified by RSD modeling. 

\subsection{Amplification of Poisson Noise in RSD Modeling}

For power spectrum, the Poisson shot noise depends on the number density $\bar{n}$. The term `shot noise' refers to the self-contribution of particles to statistical measurements, while `Poisson' indicates that the matter distribution is considered as a Poisson sampling of an underlying smooth density field. For discrete particles, the measured noisy power spectrum, denoted as $P_{\rm N}(k)$, is related to the true power spectrum $P(k)$ as \citep{Feldman1994}
\begin{equation}
P_{\rm N}(k)=P(k)+\frac{1}{\bar{n}},
\label{eq:11}
\end{equation}

The above Poisson noise effect is generic, applying to both real and redshift-space measurements. This means that both the RSD model predictions (made using measured real-space power spectra) and the redshift-space power spectra \textit{directly} measured from the simulation are affected by Poisson noise. However, the former exhibit a more pronounced effect (cf.~right panel of Fig.~\ref{fig:Kaiser1} or Fig.~\ref{fig:Kaiser2} below). The reason is as follows. Eq.~\eqref{eq:11} indicates that Poisson noise introduces an additive component to the power spectrum, while RSD, cf.~Eq.~\eqref{eq:1}, applies a multiplicative factor \citep{1992ApJ...385L...5H} to the real-space power spectrum:
\begin{equation}
    P^{\rm (S)}(k) = \left(1+\frac{2\beta}{3}+\frac{\beta^2}{5}\right)P^{\rm (R)}(k).
\end{equation}
This multiplicative factor amplifies the impact of Poisson noise on the RSD model, with
\begin{equation}\label{eq:Delta_PSN}
    \Delta = \frac{P^{\rm (S)}_{\rm model}(k)}{P^{\rm (S)}_{\rm sim}(k)} = \frac{(1+\epsilon)\left[P(k)+1/\bar{n}\right]}{(1+\epsilon)P(k)+1/\bar{n}},
\end{equation}
where we have defined
\begin{equation}
    1+\epsilon \equiv 1+\frac{2\beta}{3}+\frac{\beta^2}{5},
    \label{eq:13}
\end{equation}
and $P(k)$ is the theoretical (i.e., no Poisson noise) real-space power spectrum of the considered tracer. Using the fact that $\bar{n}P(k)\gg1$, Eq.~\eqref{eq:Delta_PSN} gives approximately
\begin{equation}
    \Delta =1+\frac{1}{\bar{n}P+1}\frac{\epsilon}{1+\epsilon} \approx 1+ \frac{1}{\bar{n}P}\frac{\epsilon}{1+\epsilon},
    \label{eq:14}
\end{equation}
which is the error, or bias, introduced in $\Delta$ by Poisson noise.

Eq.~\eqref{eq:14} demonstrates that the RSD modeling amplifies the intrinsic Poisson noise present in the real-space power spectrum. The size of the amplification depends therefore on the distortion parameter $\beta$, and hence on the tracer bias $b$ which itself varies with the tracer type, redshift and number density. In particular, even for the same number density, different tracer types have different bias values, and we expect the effect of Poisson noise to be amplified to different extents.

To examine this in more detail, we compare model performance across dark matter (sub)samples of varying number density, obtained by randomly down-sampling the simulation particles based on their IDs. In Fig.~\ref{fig:Kaiser2} we show the large-scale errors for four different dark-matter (sub)samples, using the Kaiser model. The red circles represent the case of including all particles with a number density of $8.06 \times 10^{1} (\rm Mpc/ \mathit h)^{-3}$ (i.e., the same as the red crosses in the left panel of Fig.~\ref{fig:Kaiser1}), with a $<1\%$ error on the largest scales. Even when the number density decreases to $8.06 \times 10^{-3} (\rm Mpc/ \mathit h)^{-3}$ (blue crosses), which corresponds to $10^{-4}$ of the total sample, the models still agree well with the simulation data. {Note that for downsampled particle catalogs the bias is $b=1$ which we have checked explicitly}.

However, the RSD-model-induced error in $\Delta$ becomes noticeable for sparser samples, such as those with $10^{-5}$ or $5 \times 10^{-6}$ of the total sample, as indicated by the yellow squares and cyan stars respectively. In these cases, the model performance is similar to that of halos, as shown in right panel of Fig.~\ref{fig:Kaiser1}, with errors up to a few percent, suggesting that lower number density is associated with higher amplified Poisson noise error. As an example, for $\bar{n} = 8 \times 10^{-4}(\smpc)^{-3}$, the amplified error is up to $3$\% on scales $k\simeq0.015$ -- $0.04\,h\textrm{Mpc}^{-1}$. It is expected to be stronger on even larger scales (not probed by the Millennium simulation), since $P(k)$ peaks at around $k\simeq0.015\,h\textrm{Mpc}^{-1}$.

The prediction of Eq.~\eqref{eq:14} is shown as dashed lines in Fig.~\ref{fig:Kaiser2} for the two lower-density particle catalogs. These lines roughly describe the trend of large-scale errors from simulation measurements. We also show the direct ratio of Poisson noise to power spectrum $1/\bar n P^{\rm (S)}_{\rm sim}$ as shaded regions for the two lowest-density samples. 

\begin{figure}
        \centering
	\includegraphics[width=.8\textwidth] {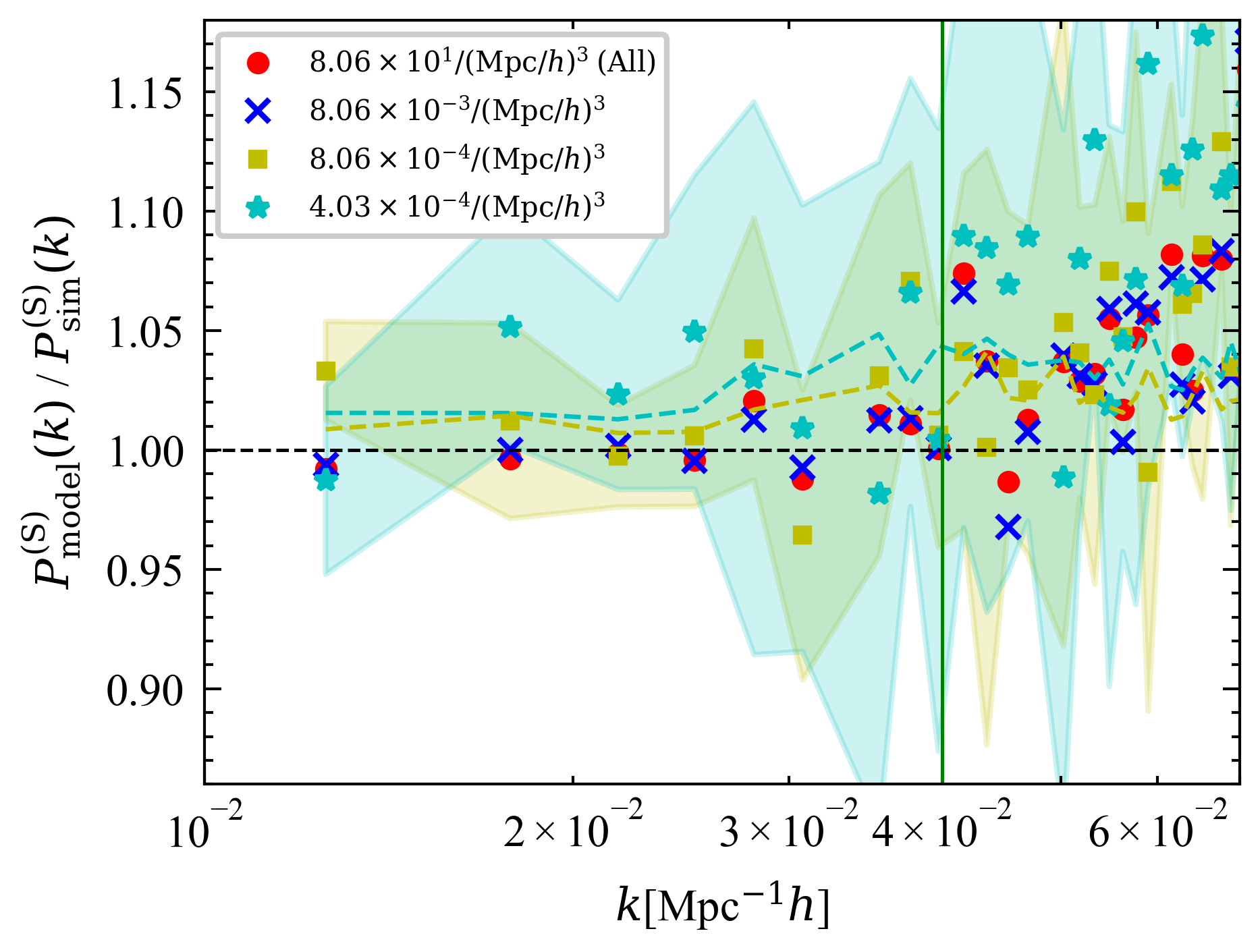}
    \caption{The large-scale RSD modeling errors and Poisson noise for dark matter particles with different number densities. Different symbols represent dark matter particle samples with different number densities. The dashed lines represent the predictions by Eq.~\eqref{eq:14} for the two lowest-density samples. The yellow and cyan shaded areas are the direct ratio of Poisson noise to power spectrum $1/(\bar n P^{\rm (S)}_{\rm sim})$ for the two lowest-density samples.} 
    \label{fig:Kaiser2}
\end{figure}

We note again that this error is generated because we have used the measured real-space power spectra of a tracer catalog to predict the corresponding redshift-space spectra. If we start from theoretical predictions of the real-space spectra \citep[e.g.,][]{Matsubara2008,Lawrence2010,Lawrence2017}, based on perturbation theory or emulator, there should be no issues.

\subsection{Numerical Verification with Different Tracers}

\begin{figure*}
    \includegraphics[width=\textwidth] {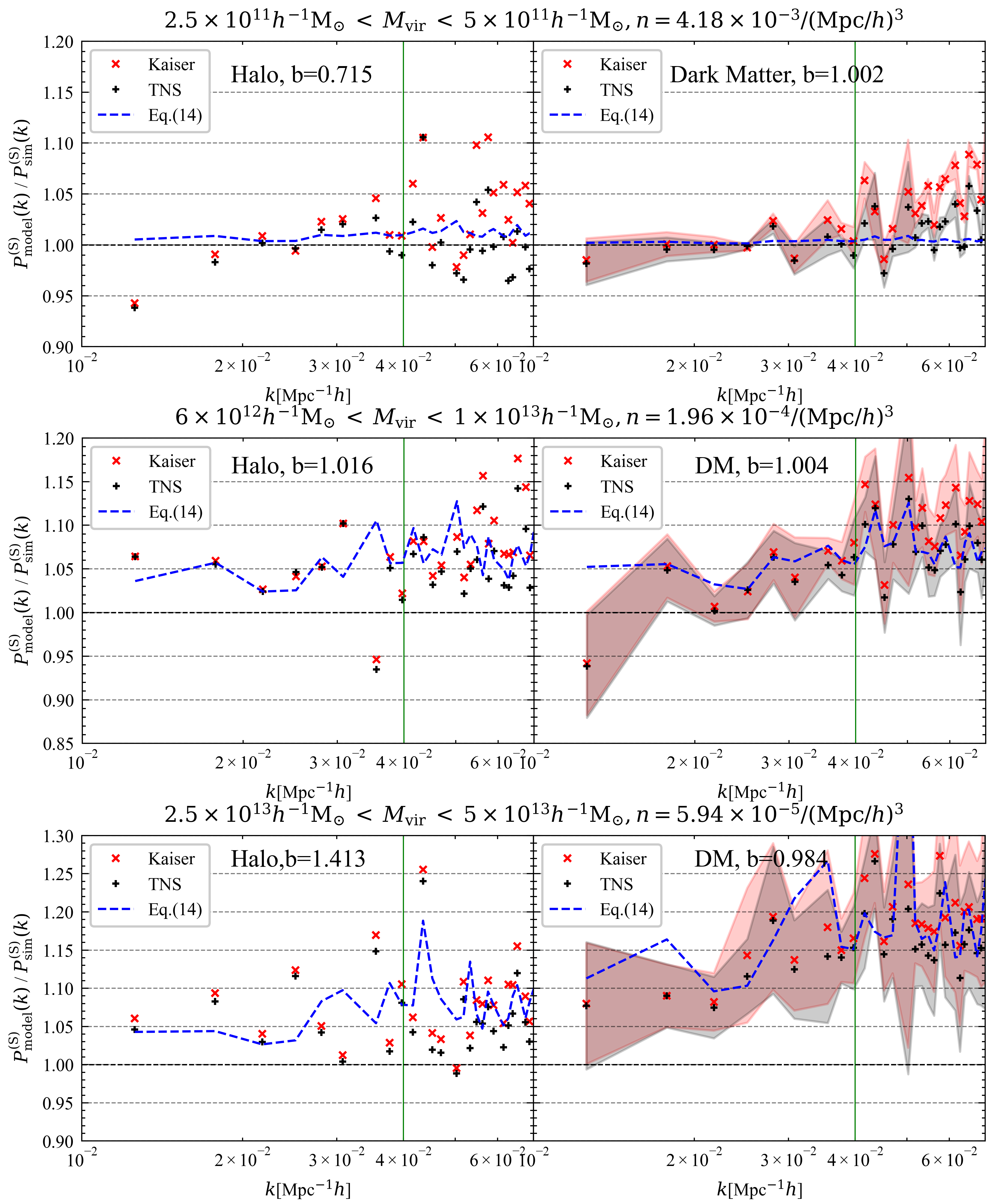}
    \caption{The large-scale errors in the RSD modeling, for halos (left column) and dark matter particles (right panel) with the equal number densities. This figure differs from Fig.~\ref{fig:Kaiser1} in that here the number densities of the left and right panels are equal in each row. For each specified number density, eight downsamples of dark matter particles are generated, with their mean and variance (shown as a shaded area) plotted in each right sub-figure. The blue dashed line shows the prediction by Eq.~\eqref{eq:14}.}
    \label{fig:Kaiser3}
\end{figure*}

\begin{figure*}
	\includegraphics[width=\textwidth] {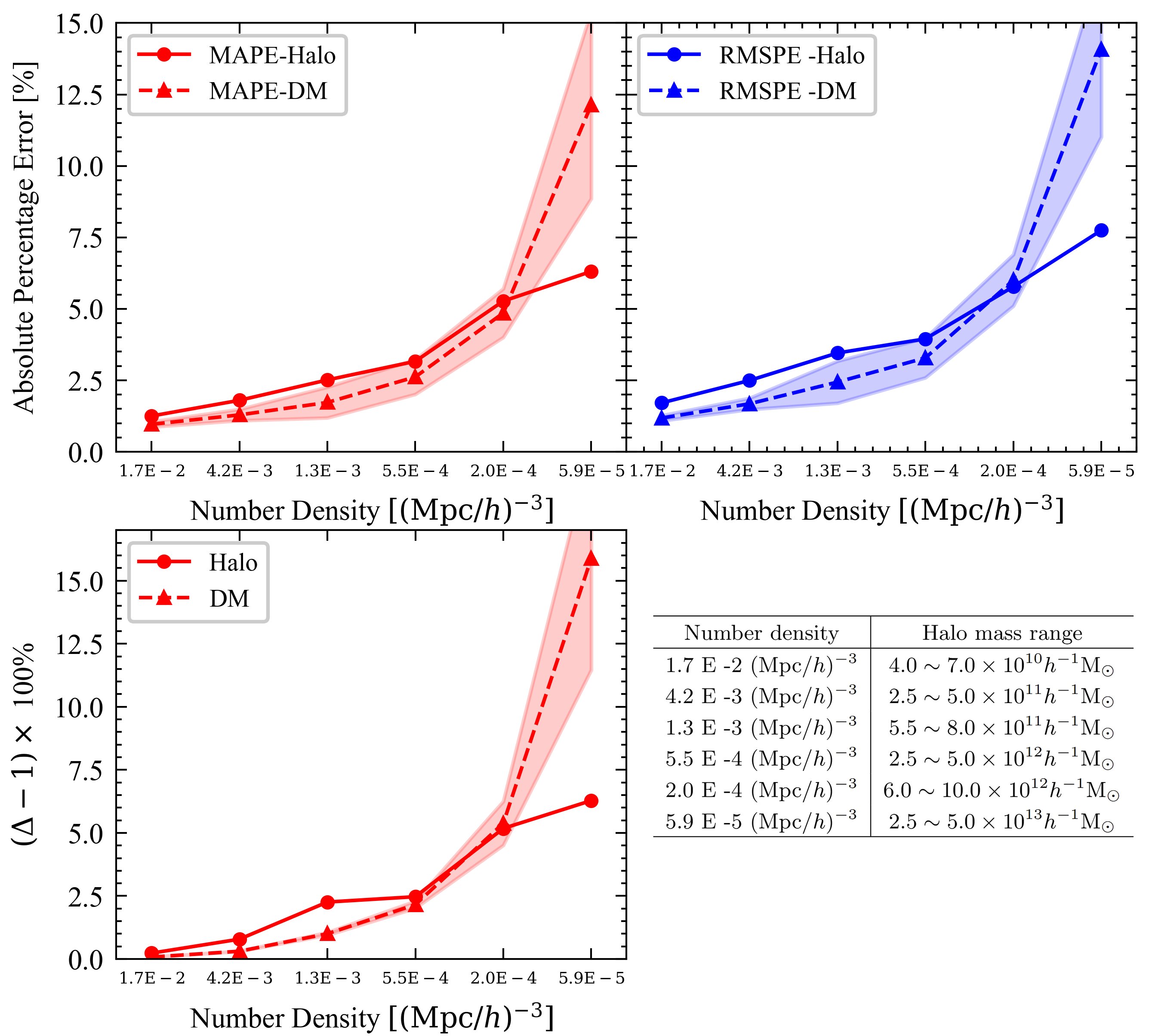}
    \caption{The Mean Absolute Percentage Error (MAPE; upper left) and Root Mean Square Percentage Error (RMSPE; upper right) of halo and dark matter particles with equal number density (shown in the horizon axes). These can be compared with the theoretical prediction of Eq.~\eqref{eq:14} (lower left) of $\Delta$ averaged over scales $k<$ 0.04 $\kmpc$. The circles and triangle respectively represent the results for halos and dark matter particles. The shaded areas denote the $1$-$\sigma$ uncertainties measured from 8 random subsamples of dark matter particles. The halo mass ranges corresponding to the chosen number densities are shown in the table on the lower right. At $\bar{n} \lesssim 2\times10^{-4}({\rm Mpc}/h)^{-3}$, the halo errors exceed those of the dark matter sample, consistent with the theoretical predictions.}
    \label{fig:fig4_shot}
\end{figure*}

The validity of Eq.~\eqref{eq:14} can be further verified using different tracer typles with the same number density but different values of the bias $b$, which the $\beta$ parameter in Eq.~\eqref{eq:13} and Eq.~\eqref{eq:14} depends on. This is done in Fig.~\ref{fig:Kaiser3} and Fig.~\ref{fig:fig4_shot}, 
in which we have chosen a few halo catalogs and downsampled the dark matter particles to match their number densities.

In Fig.~\ref{fig:Kaiser3}, we show this for 3 halo samples respectively in the mass range $[2.5 \times 10^{11} h^{-1} \mathrm{M}_{\odot}, 5 \times 10^{11} h^{-1} \mathrm{M}_{\odot}]$, $[6 \times 10^{12} h^{-1} \mathrm{M}_{\odot}, 1 \times 10^{13} h^{-1} \mathrm{M}_{\odot}]$, and $[2.5 \times 10^{13} h^{-1} \mathrm{M}_{\odot}, 5 \times 10^{13} h^{-1} \mathrm{M}_{\odot}]$, with number densities $4.18 \times 10^{-3}$, $1.96 \times 10^{-4}$ and $5.94 \times 10^{-5}(\rm{Mpc}$$/h)^{-3}$. The large-scale errors associated with these halo catalogs are shown in the left panels. For comparison, the right panels show the corresponding results for the downsampled particle catalogs with the same number densities. For each number density, eight downsamples of dark matter particles are generated, with their mean and variance (shown as a shaded area) plotted. For halos, all halos in a given mass bin are used, hence multiple sampling is not conducted. The blue dashed line represent the error induced by Poisson noise as predicted by Eq.~\eqref{eq:14}.

In the upper panels, the bias of halos is $b_{\rm h}=0.715$ and the average bias of dark matter samples is $b_{\rm dm}=1.002$. In the middle panels, we have $b_{\rm h}\simeq b_{\rm dm}\simeq 1$, while in the lower panels $b_{\rm h}=1.413$ and $b_{\rm dm}=0.984$. In all six cases the actual bias values are used in the predictions of Eq.~\eqref{eq:14}, and we find good agreements between these predictions and the simulation measurements. We note that, for the same number density, a lower bias results in a higher $\epsilon$ in Eq.~\eqref{eq:13} and hence greater amplification factor for the Poisson noise. This is also confirmed by comparing the left and right panels of Fig.~\ref{fig:Kaiser3}.



In order to better quantify the large-scale errors of halo and dark matter, we define the Mean Absolute Percentage Error (MAPE) and Root Mean Square Percentage Error (RMSPE) as
\begin{eqnarray}
 \rm{MAPE} &=& \rm{mean}\left[\left| \Delta(\textit k)-1 \right|\right]\times 100\%;\nonumber\\
\rm{RMSPE} &=& \sqrt{\rm{mean}\left[(\Delta(\textit k)-1 )^{2}\right]}\times 100\%,
\end{eqnarray} 
where the mean is taken for all Fourier modes at $k<0.04 \kmpc$.

In Fig.~\ref{fig:fig4_shot}, the upper panels show the MAPE and RMSPE results of halo (solid lines with circles) and downsampled dark matter particles (dashed lines with triangles) with equal number density, while the lower-left panel shows the predictions for halos and dark matter by Eq.~\eqref{eq:14} averaged over $k < 0.04\kmpc$. The theoretical predictions use the actual measurement of bias values. The shaded regions again indicate the $1$-$\sigma$ range derived from 8 samples of dark matter particles.

As the halo mass increases and the halo number density decreases, which also lead to an increase in halo bias, the MAPE and RMSPE values increase significantly. At $\bar{n} \lesssim 2\times10^{-4}({\rm Mpc}/h)^{-3}$, the halos' MAPE and RMSPE exceed those for the dark matter samples. At this crossover we indeed have $b_{\rm h}\simeq b_{\rm dm}$, which induces similarly amplified Poisson noise in RSD modeling, as can be seen by comparing the top panels of Fig.~\ref{fig:Kaiser3}. These values also agree very well with the prediction of Eq.~\eqref{eq:14}, as a further verification of the latter. 

Overall, the results show that, when the number density decreases and shot noise increases, the Poisson noise amplification becomes more significant. For fixed shot noise levels, selecting tracers with higher bias can reduce the degree of Poisson noise amplification.

\section{Conclusion and Discussion}
\label{sec:conclusion}

This study examines large-scale RSD modeling errors, evaluating the Kaiser linear model and the TNS nonlinear model. We found that RSD modeling consistently amplifies intrinsic Poisson noise from the real-space power spectrum when converting it to redshift space. This large-scale error stays well below $1\%$ with high sample densities, but can reach over $5\%$ at densities near $10^{-4}({\rm Mpc}/h)^{-3}$. Using dark matter and halo samples with different number densities, we show that the Poisson noise amplification is well-described by Eq.~\eqref{eq:14}.


As mentioned above, the error in $\Delta$ induced by RSD modeling is a result of the Poisson noise that already exists in the real-space spectrum. In many previous studies, the real-space galaxy power spectrum is typically predicted using perturbation theory, which is free from Poisson noise. In these cases, the amplified Poisson noise error we discussed above does not arise.

For emulators trained using halo or mock galaxy data, the predicted galaxy power spectrum may contain shot noise. Some of these studies \citep[e.g.,][]{Kobayashi2020,Wang2023} predict the RSD power spectrum directly from redshift-space simulation data, where the authors subtract the shot noise contribution ($1/\bar{n}$ term) {\it a priori}. There is also no need to incorporate additional RSD modeling process. {The other case is emulators that predicts real-space galaxy power spectra with shot noise contributions \citep{Donald2022}. When these emulators are used to predict the real-space power spectra for RSD modeling, it is advisable to have an extra step of shot noise subtraction, for reasons discussed here.}

{Another implication of this study is that general modeling techniques based on discrete density fields derived from tracers—such as RSD modeling or BAO reconstruction—may either enhance or reduce the intrinsic shot noise component. Specifically, as shown earlier, transforming from the real-space density field to redshift space increases shot noise, whereas converting from redshift space to real space reduces it. For the case of BAO reconstruction, we briefly discuss the process here. The displacement field $\Psi$ is computed from the density field:}
\begin{equation}
\Psi(\boldsymbol{k})=i \frac{\boldsymbol{k}}{k^2} \delta_{\text {obs }}(k)=i \frac{\boldsymbol{k}}{k^2}\left(\delta_{\text {true }}(k)+\delta_{\text {shot }}(k)\right),
\end{equation}
{in which $\Psi_{\text {shot }}(\boldsymbol{k})\equiv i \frac{\boldsymbol{k}}{k^2} \delta_{\text {shot }}(k)$ is the error term in the displacement field arising from shot noise. The power spectrum of its component along the x-direction can be obtained by integrating over the power spectrum:}
\begin{equation}
\begin{aligned}
P_{\Psi_{\text {shot},x}}(k)&=\frac{k_x^2}{k^4} P_{\delta_ \text {shot}}=\frac{k_x^2}{k^4} \frac{1}{\bar{n}} \cdot|W(k)|^2\\
&= \frac{1}{3k^2}\frac{1}{\bar{n}} \exp \left(-k^2 R_{\mathrm{sm}}^2\right),
\end{aligned}
\end{equation}
and its variance is given by
\begin{equation}
\begin{aligned}
\sigma_{\Psi_{\text {shot},x}}^2&=\int \frac{d^3 k}{(2 \pi)^3} P_{\Psi_{\text {shot}, x}}(k) \\
&=\int_0^{\infty} \frac{k^2 d k}{2 \pi^2} \cdot \frac{1}{3 k^2 \bar{n}}\exp \left(-k^2 R_{\mathrm{sm}}^2\right)\\
&=\frac{1}{12 \pi^{3 / 2} \bar{n} R_{\text {sm }}},
\end{aligned}
\end{equation}
{where $R_{\text {sm }}$ is the smoothing scale adopted. Therefore, the standard deviation of the shot-noise term can be approximated as:}

\begin{equation}
\sigma_{\Psi_{\text {shot}}}\approx\sqrt{3\sigma_{\Psi_{\text {shot},x}}^2}=\frac{(\bar{n}R_{\text {sm }})^{-1 / 2}}{2 \pi^{3/4} } =\frac{0.211}{\sqrt{\bar{n}R_{\text {sm }}}}
\end{equation}

{For the galaxies with  a number density of around $5\times10^{-4}({\rm Mpc}/h)^{-3}$ and the BAO smoothing scale $R_{\text {sm }}=10 {\rm 
Mpc}/h $ , the standard deviation of the shot-noise term in the displacement field is about 3 ${\rm Mpc}/h$. }

{Similarly, we can estimate the variance of the displacement field using the matter power spectrum.}
\begin{equation}
\begin{aligned}
\sigma_{\Psi_x}^2&=\int \frac{d^3 k}{(2 \pi)^3} P_{\Psi_x}(k) \\
&=\int_0^{\infty} \frac{k^2 d k}{2 \pi^2} \cdot \frac{1}{3k^2}P_{\delta\delta}(k)\exp \left(-k^2 R_{\mathrm{sm}}^2\right)\\
&\approx \frac{1200}{6\pi^2},
\end{aligned}
\end{equation}
{so that $\sigma_{\Psi}=\sqrt{3\sigma_{\Psi}^2}\approx8 {\rm Mpc}/h$. We have  directly measured $\sigma_\Psi$ using the Millennium Simulation data at $z=0$ and found a result of $\approx7.5 {\rm Mpc}/h$, which aligns well with the above estimate. Consequently, shot noise can introduce a random error exceeding one-third in the displacement field. This may impact the sharpening efficacy of BAO peak reconstruction (though a more quantitative study of this effect lies beyond the scope of this paper). Increasing the number density and fine-tuning the smoothing scale can substantially reduce this error source.}



\acknowledgments

We thank Tianxiang Mao, Yanchuan Cai and Pengjie Zhang for very useful discussions. This work is supported by the National Key R\&D Program of China ( 2022YFA1602901), the NSFC grant (Nos. 11988101, 11873051, 12125302, 11903043) and CAS Project for Young Scientists in Basic Research Grant (No. YSBR-062).






\bibliographystyle{JHEP}
\bibliography{biblio.bib}

\end{document}